\documentclass{article}

\usepackage{arxiv}

\usepackage[utf8x]{inputenc} 
\usepackage[T1]{fontenc}    
\usepackage{hyperref}       
\usepackage{url}            
\usepackage{booktabs}       
\usepackage{amsfonts}       
\usepackage{nicefrac}       
\usepackage{microtype}      
\usepackage{lipsum}

\title{On the probability flow in the Stock market I: The Black-Scholes case}

\author{
 Ivan Arraut\\
  Lee Shau Kee School of Business and Administration\\
 The Open University of Hong Kong,\\
  30 Good Shepherd Street, Homantin, Kowloon \\
  \texttt{ivanarraut05@gmail.com} \\
   \And
 Alan Au \\
   Lee Shau Kee School of Business and Administration\\
 The Open University of Hong Kong,\\
  30 Good Shepherd Street, Homantin, Kowloon \\
  \texttt{kmau@ouhk.edu.hk} \\
  \AND
   Alan Ching-biu Tse \\
 Department of Marketing, The Chinese University of Hong Kong, \\
  Cheng Yu Tung Building 12 Chak Cheung Street Shatin, N.T., Hong Kong \\\\
  Lee Shau Kee School of Business and Administration\\
 The Open University of Hong Kong,\\
  30 Good Shepherd Street, Homantin, Kowloon\\
 \And
  Jo$\tilde{\bf a}$o Alexandre Lobo Marques \\
  SBU, University of Saint Joseph\\
  Estrada Marginal da Ilha Verde, 14-17, Macao, China\\
}

\begin{document}
\maketitle

\begin{abstract}
It is known that the probability is not a conserved quantity in the stock market, given the fact that it corresponds to an open system. In this paper we analyze the flow of probability in this system by expressing the ideal Black-Scholes equation in the Hamiltonian form. We then analyze how the non-conservation of probability affects the stability of the prices of the Stocks. Finally, we find the conditions under which the probability might be conserved in the market, challenging in this way the non-Hermitian nature of the Black-Scholes Hamiltonian.   
\end{abstract}

\keywords{First keyword \and Second keyword \and More}

\section{Introduction}

The financial equations, when expressed in the eigenvalue form, are represented by non-Hermitian Hamiltonian operators \cite{1, 2, 3}. Non-Hermitian Hamiltonians are considered to be pathological in ordinary Quantum Mechanics because they are connected with non-conservation of the information (probability) \cite{4}. In fact, unitarity (conservation of probability) is considered to be fundamental for a consistent formulation of Quantum Mechanics. However, there are some situations where it is interesting to analyze possible loses in information in Quantum systems or non-conservation of probability \cite{5, 6}. This comes out to be relevant when we analyze open systems \cite{6}. Indeed, it is understood that the Stock market itself is an open system \cite{5}. There are some situations where the unitarity can be restored, even if the Hamiltonian is non-Hermitian. This happens when the system under analysis respects the combined symmetries of Parity-Time Reversal (PT) \cite{UR}. Some consistent physical systems, cannot avoid the loss of information or loss of unitarity naturally. This is the case of Black-Holes where the Hawking radiation comes out to be connected with the loss of unitarity in a physical system \cite{Hawking}. Interesting practical applications of the Hawking radiation effect can be found in neural networks and Deep Learning \cite{Hawking2}. In this paper, we will concentrate on the Hamiltonian related to the Black-Scholes (BS) equation taken here as the ideal case with constant volatility \cite{1}. The BS Hamiltonian is non-Hermitian, implying a possible non-conservation of probability. We will obtain the probability currents for this case and we will study the continuity equation in both cases. We will also explain the effects of the non-conservation of probability over the prices of the Stocks and we will derive the conditions under which the information might be preserved in the market. The paper is organized as follows: In Sec. (\ref{CPQM}), we will review the standard conservation of probability in Quantum Mechanics (QM). In Sec. (\ref{BSHAM}), we review the the derivation of the BS equation together with its Hamiltonian form. In Sec. (\ref{Evolprob}), we analyze the flow of probability in the BS equation by looking at the Kernel of the solution and analyzing the continuity equation. We then find the conditions under which the probability might be conserved. In Sec. (\ref{Section5}), we explain how the flow of probability is connected with the observed prices in the market. Finally, in Sec. (\ref{Conclu}), we conclude.   

\section{Conservation of probability in Quantum Mechanics}   \label{CPQM}

The Schr\"odinger equation in Quantum Mechanics is defined as \cite{4}

\begin{equation}   \label{eq1}
\hat{H}\psi({\bf r}, t)=E\psi({\bf r}, t).  
\end{equation}
Here the Hamiltonian $\hat{H}$ is an function operator, depending mainly on the momentum-operator $\hat{{\bf p}}$ and on the position operator $\hat{\bf r}$. The way how the operators act over the wave-function $\psi$ defined in eq. (\ref{eq1}), depends on the space where they are defined. In position space, we have a corresponding wave-function defined as $\psi({\bf r}, t)$ and the position operator acts as a simple product $\hat{\bf r}\psi({\bf r}, t)={\bf r}\psi({\bf r}, t)$. The momentum operator on the other hand, acts as a derivative operator defined as $\hat{\bf p}\psi({\bf r}, t)=-i\hbar{\bf \nabla}_{\bf r}\psi({\bf r}, t)$. The solution of eq. (\ref{eq1}) is defined by the complex wave-function $\psi({\bf r}, t)$. The probability amplitude for some outcome for a series of eigenvalues is in general defined by the square of the amplitude of the wave-function as

\begin{equation}
P=\int_V d^3x \vert\psi({\bf r}, t)\vert^2. 
\end{equation}
The integral is evaluated over the volume $V$ under analysis. Note that since the sum of probabilities has to be equal to one, then the previous integral evaluated over the whole space is equal to one. Given the definition of probability, it comes out to be conserved if the evolution of the system is unitary. Unitary evolution demands the norm $\vert\psi({\bf r}, t)\vert^2$ to be conserved in time. Then the time evolution of the system is defined by an unitary operator $\hat{U}=e^{-i\hat{H}t}$, where $\hat{H}$ is the Hamiltonian of the system. Note that the operator $\hat{U}$ is unitary if the Hamiltonian is Hermitian. 

\subsection{Conservation of the norm $\vert\psi({\bf r}, t)\vert$}

We can find the evolution of the norm in general by calculating the time-derivative of $\vert\psi({\bf r}, t)\vert^2$ as follows

\begin{equation}   \label{derivation}
\frac{\partial \vert\psi({\bf r}, t)\vert^2}{\partial t}=\psi^*\frac{\partial\psi}{\partial t}+\frac{\partial\psi^*}{\partial t}\psi.    
\end{equation}
Here we will denote $\rho({\bf r}, t)=\vert\psi({\bf r}, t)\vert^2$. We can consider the Hermitian Hamiltonian operator as 

\begin{equation}   \label{Hamilton}
\hat{H}=-\frac{\hbar^2}{2m}\nabla_{\bf r}^2+\hat{V}.    
\end{equation}
Here $\nabla^2={\bf\nabla}\cdot{\bf\nabla}$ and $\hat{V}$ is the potential operator which only depends on the position operator for the conservative cases. In the position space, the operator $\hat{V}$ acts as a simple function. Considering the Schr\"odinger equation (\ref{eq1}), together with the Hamiltonian (\ref{Hamilton}), inside the derivation (\ref{derivation}), then we get

\begin{equation}   \label{everytime}
\frac{\partial\rho}{\partial t}=\frac{\partial}{\partial x}\left(\frac{i\hbar}{2m}\left(\psi^*\frac{\partial\psi}{\partial x}-\frac{\partial\psi^*}{\partial x}\psi\right)\right),   
\end{equation}
where we have simplified the problem to one-dimension without loss of generality. The spatial integration of eq. (\ref{everytime}) in both sides gives us 

\begin{equation}   \label{dt}
\frac{d}{dt}\int_{-\infty}^\infty\rho(x, t)dx=\frac{i\hbar}{2m}\left(\psi^*\frac{\partial\psi}{\partial x}-\frac{\partial\psi^*}{\partial x}\psi\right)^{\;\;\infty}_{-\infty}. 
\end{equation}
Then we can perceive that there exists two different conditions which guarantee that the probability is conserved. The first one is the vanishing of the probability current defined as

\begin{equation}   \label{current}
-{\bf j}=\frac{i\hbar}{2m}\left(\psi^*\frac{\partial\psi}{\partial x}-\frac{\partial\psi^*}{\partial x}\psi\right),
\end{equation}
due to the equality 

\begin{equation}
\psi^*\frac{\partial\psi}{\partial x}=\frac{\partial\psi^*}{\partial x}\psi.   
\end{equation} 
This happens for example for a particle confined between two walls (infinite potential-well) \cite{4}. Here the selected sign in eq. (\ref{current}) is a matter of convention. Note that when ${\bf j}=0$, then the probability is conserved in eq. (\ref{dt}), no matter what other characteristics the wave-function $\psi({\bf r}, t)$ has. The second condition for the probability to be conserved in time is the asymptotic vanishing of the wave-function $\psi({\bf r}, t)$ in the limit ${\bf r}\to\infty$. This previous analysis is valid in ordinary Quantum Mechanics. However, when we go to the Financial world, some variations specifically related to the time-evolution of the system has to be done. We will come back to this point later. at this point it is convenient to define the continuity equation, imagining that the probability is a fluid moving through the volume under analysis. The continuity equation is defined as

\begin{equation}
\frac{\partial\rho}{\partial t}+{\bf \nabla}\cdot{\bf j}=0.    
\end{equation}
This equation tells us that the change in time of the probability of the system, depends on the current flow of probability. If the system is isolated (no-boudaries or boundaries located at the infinity), then there will not be flow of information through the boundaries and then the probability is conserved. 

\section{Financial Hamiltonians: The Black-Scholes case}   \label{BSHAM}

Here we consider the stock price as $S(t)$, which is normally taken as a random stochastic variable evolving in agreement to a stochastic differential equation given by

\begin{equation}   \label{rainbowl}
\frac{dS(t)}{dt}=\phi S(t)+\sigma SR(t).
\end{equation}   
Here $\phi$ is the expected return of the security, $R(t)$ is the Gaussian white noise with zero mean and $\sigma$ is the volatility \cite{1}. The fundamental analysis of Black and Scholes, exclude the volatility such that we can guarantee the evolution of the price of the stock with certainty \cite{9}. In this way, by imposing $\sigma=0$, we obtain a simple solution for the equation (\ref{rainbowl}) as

\begin{equation}
S(t)=e^{\phi t}S(0).    
\end{equation}
The possibility of arbitrage is excluded if we can make a perfect hedged portfolio. In this sense, any possibility of uncertainty is excluded and we can analyze the evolution of the price free of any white noise \cite{1}. We can consider the following portfolio

\begin{equation}   \label{forever}
\Pi=C-\frac{\partial C}{\partial S}S.
\end{equation}
This is a portfolio where an investor holds the option and then {\it short sells} the amount $\frac{\partial \psi}{\partial S}$ for the security $S$. By using the Ito calculus (stochastic calculus) \cite{1}, it is possible to demonstrate that

\begin{equation}   \label{BS}
\frac{d\Pi}{dt}=\frac{\partial C}{\partial t}+\frac{1}{2}\sigma^2S^2\frac{\partial^2 C}{\partial S^2}.
\end{equation}
Here the change in the value of $\Pi$ does not have any uncertainty associated to it \cite{1}. The random term has disappeared due to the choice of portfolio. Since here we have a risk-free rate of return for this case (no arbitrage) \cite{Europe, Epjons}, then the following equation is satisfied

\begin{equation}   \label{hedged}
\frac{d\Pi}{dt}=r\Pi.
\end{equation} 
If we use the results (\ref{forever}) and (\ref{BS}), together with the previous equation, then we get

\begin{equation}   \label{BSeq}
\frac{\partial C}{\partial t}+rS\frac{\partial C}{\partial S}+\frac{1}{2}\sigma^2S^2\frac{\partial^2C}{\partial S^2}=rC.
\end{equation}
This is the Black-Scholes equation \cite{Op2, Op3, Merton2}, which is independent of the expectations of the investors, defined by the parameter $\phi$, which appears in eq. (\ref{rainbowl}). In other words, in the Black-Scholes equation, the security (derivative) price is based on a risk-free process. The Basic Assumptions of the Black-Scholes equation are:\\
1). The spot interest rate $r$ is constant.\\
2). In order to create the hedged portfolio $\Pi$, the stock is infinitely divisible, and in addition it is possible to short sell the stock.\\
3). The portfolio satisfies the no-arbitrage condition.\\
4). The portfolio $\Pi$ can be re-balanced continuously.\\
5). There is no fee for transaction.\\ 
6). The stock price has a continuous evolution. \\

\subsection{Black-Scholes Hamiltonian formulation}

We will explain how the eq. (\ref{BSeq}), can be expressed as an eigenvalue problem after a change of variable. The resulting equation will be the Schr\"odinger equation with a non-Hermitian Hamiltonian. For eq. (\ref{BSeq}), consider the change of variable $S=e^x$, where $-\infty<x<\infty$. In this way, the BS equation becomes

\begin{equation}   \label{BSHamiltonian2}
\frac{\partial C}{\partial t}=\hat{H}_{BS}C,
\end{equation} 
where we have defined the operator  

\begin{equation}   \label{BSHamiltonian}
\hat{H}_{BS}=-\frac{\sigma^2}{2}\frac{\partial^2}{\partial x^2}+\left(\frac{1}{2}\sigma^2-r\right)\frac{\partial}{\partial x}+r.
\end{equation}
as the BS Hamiltonian. Note that the resulting Hamiltonian is non-Hermitian since $\hat{H}\neq\hat{H}^+$ \cite{non-her}. This will be an important point to analyze in the light of the probability conservation. In addition, note that since the spot interest rate $r$ is constant, then the potential term is just a constant term. This means that the vacuum condition is trivial for this case. Under the BS Hamiltonian, the evolution in time of the Option is non-unitary in general (in addition, the Hamiltonian non-necessarily obeys the $PT$ symmetry). This means that the probability is not necessarily preserved in time, although it is certainly well-defined and its total value is equal to one, if the normalization factor is updated time to time. In general, as it was mentioned in the introduction, there are some cases in ordinary Quantum Mechanics, as well as in Quantum Field Theory, where it is interesting to explore non-Hermitian Hamiltonians (Lagrangians) \cite{5, 6, UR}. Based on the previous explanations, we cannot expect the financial market to obey unitarity. This is precicely the point which we want to analyze in a deep detail in this paper.  

\section{The evolution of probability in the Black-Scholes equation}   \label{Evolprob}

The solutions for the BH equation is well-known \cite{1}. In this section we will explore the Kernel for the evolution of the prices in the stock market. This quantity depends on the Hamiltonian under analysis. For understanding this point further, we can generalize the equation (\ref{BSHamiltonian2}) to any Hamiltonian $\hat{H}$, obtaining then the corresponding solution

\begin{equation}
C(t, x)=e^{iEt}C(0, x).    
\end{equation}
Here $E$ is the eigenvalue for the Hamiltonian $\hat{H}$. The same result can be expressed in the bra-ket notation as

\begin{equation}
\vert C, t>=e^{Et}\vert C, 0>.    
\end{equation}
Considering the final condition at $t=T$, we have

\begin{equation}
\vert C, T>=e^{ET}\vert C, 0>=g,    
\end{equation}
where $g$ is just the pay-off function. The initial stock price is then defined as

\begin{equation}
\vert C, 0>=e^{-ET}g,
\end{equation}
and then 

\begin{equation}
\vert C, t>=e^{-E\tau}g,    
\end{equation}
with the time $\tau=T-t$ running backwards. Note that the previous solution represents an exponential decaying behavior. Projecting the previous result over $<x\vert$, implies

\begin{equation}
C(x, t)=<x\vert C, t>=<x\vert e^{-\tau H}\vert g>=\int^\infty_{-\infty}dx'<x\vert e^{-\tau H}\vert x'>g(x').    
\end{equation}
Note that in the last step we have used the completeness relation $\int^\infty_{-\infty} dx'\vert x'><x'\vert=\hat{I}$. Then we define the pricing Kernel as

\begin{equation}   \label{KERNEL}
p(x, \tau, ; x')=<x\vert e^{-\tau \hat{H}}\vert x'>.    
\end{equation}
\subsection{Black-Scholes pricing Kernel}

By introducing the Hamiltonian (\ref{BSHamiltonian}) inside eq. (\ref{KERNEL}), we should be able to find the explicit form of the pricing Kernel for the BS case. Before that, we have to work-out a few definitions. We start with

\begin{equation}
\delta(x-x')=<x\vert x'>=\int^\infty_{-\infty}\frac{dp}{2\pi}e^{ip(x-x')}=\int^\infty_{-\infty}\frac{dp}{2\pi}<x\vert p><p \vert x'>.    
\end{equation}
Then the completeness relation 

\begin{equation}
\int^\infty_{-\infty}\frac{dp}{2\pi}\vert p><p \vert>=\hat{I},    
\end{equation}
is satisfied. In addition, the previous results define the product

\begin{equation}
<x\vert p>=e^{ipx}, \;\;\;\;\;<p\vert x>=e^{-ipx}    
\end{equation}
The BS Kernel can be obtained by using eq. (\ref{KERNEL}) with $\hat{H}=\hat{H}_{BS}$. For that purpose, it is convenient to re-express the same equation as

\begin{equation}   \label{propa}
p_{BS}(x, \tau, ; x')=\int^\infty_{-\infty}\frac{dp}{2\pi}<x\vert e^{-\tau \hat{H}_{BS}}\vert p><p\vert x'>.   
\end{equation}
For evaluating explicitly this equation, we have to use the following result

\begin{equation}
<x\vert\hat{H}_{BS}\vert p>=E_{BS}<x\vert p>=E_{BS}e^{ipx}.    
\end{equation}
Here $E_{BS}$ is the eigenvalue of the Hamiltonian operator corresponding to the state $\vert p>$. In order to find it explicitly, we need to evaluate the following matrix element

\begin{equation}   \label{Hammatrix}
<p\vert\hat{H}_{BS}\vert x>=<x\vert\hat{H}^+_{BS}\vert p>^*=\left(\frac{1}{2}\sigma^2p^2+i\left(\frac{1}{2}\sigma^2-r\right)p+r\right)e^{-ipx}.    
\end{equation}
Note that here we are using $\hat{p}\vert p>=p\vert p>$. This means that the operator $\hat{H}_{BS}$ acts over the momentum space. Remember that in position space, the momentum operator is a derivative operator. If we use the result (\ref{Hammatrix}), then the Kernel defined by eq. (\ref{propa}) becomes

\begin{equation}   \label{Hammatrixagain}
p_{BS}(x, \tau, ; x')=e^{-r\tau}\frac{1}{\sqrt{2\pi\tau\sigma^2}}e^{-\frac{1}{2\tau \sigma^2}\left(x-x'+\tau\left(r-\frac{\sigma^2}{2}\right)\right)^2}.    
\end{equation}
Here $\tau=T-t$. Note that since the Kernel contains all the functional dependence of the states under analysis, the conservation of probability comes out to be equivalent to some conditions imposed over the Kernel. First we can evaluate the time-derivative of the Kernel as

\begin{equation}
\frac{\partial p_{BS}}{\partial t}=\frac{\partial p_{BS}}{\partial \tau}\frac{\partial \tau}{\partial t}.    
\end{equation}
Since $\partial\tau/\partial t=-1$, then we get

\begin{equation}
\frac{\partial p_{BS}}{\partial t}=-\frac{\partial p_{BS}}{\partial \tau}\neq0,    
\end{equation}
in general. The previous derivative is equal to zero for some relations between the variables and between parameters. In fact we obtain

\begin{equation}   \label{rate-free}
r=-\frac{\sigma^2}{2},\;\;\;\;\;if \;\;\;\;\;\frac{\partial p_{BS}}{\partial t}=0,\;\;\;\tau\to\infty.    
\end{equation}
we also have the additional condition

\begin{equation}
\tau=\frac{(x-x')^2}{\sigma^2},\;\;\;\;\;if \;\;\;\;\;\frac{\partial p_{BS}}{\partial t}=0,\;\;\;\tau\to0,\;\;\;(arbitrary\;\;r)
\end{equation}
Note that in eq. (\ref{rate-free}), the risk-free interest rate takes negative values under the imposed limit. In fact, $r$ can be either positive or negative \cite{pn}. However, in order to satisfy the demands of the BS equation, we impose the condition $r\geq0$, avoiding then negative risk-free interest rates. Since eq. (\ref{rate-free}) allows the possibility of having negative values for $r$, there must be a limit value for the time $\tau$ where $r=0$. This important limit can be found if we impose $r\to0$ in the time-derivative of the Kernel (\ref{Hammatrixagain}). Assuming again conservation of probability ($\frac{\partial p_{BS}}{\partial t}=0$), we get the polynomial 

\begin{equation}
\tau^2+\frac{4}{\sigma^2}\tau-4\frac{(x-x')^2}{\sigma^4}=0. 
\end{equation}
If we solve for $\tau$, then we get

\begin{equation}
\tau_{r=0}=-\frac{2}{\sigma^2}\left(1\pm\sqrt{1+(x-x')^2}\right).    
\end{equation}
Here the lower sign is the correct one in order to keep $\tau$ positive. The limit value $\tau^{r=0}_{min}=0$ ($t=t_i$, initial value) corresponds to the condition $x\to x'$, which is consistent with the fact the $x'$ corresponds to the initial price. Note that when $x>>x'$, namely, when $(x-x')\to\infty$, then we have the result 

\begin{equation}
\tau^{r=0}_{max}\approx \frac{2\vert x-x'\vert}{\sigma^2}.    
\end{equation}
Other important limit to be considered is $r=\frac{\sigma^2}{2}$. For this limit, the condition $\frac{\partial p_{BS}}{\partial t}=0$ gives

\begin{equation}   \label{upper}
\tau=-\frac{1}{4r}\left(1\pm\sqrt{1+\frac{8r(x-x')^2}{\sigma^2}}\right).  
\end{equation}
Since we know again that the condition $\tau\geq0$ must be satisfied, then we select the lower sign in eq. (\ref{upper}). Note that for the minimal value of $\tau^{r=\sigma^2/2}_{min}=0$, we get

\begin{equation}
\frac{r(x-x')^2}{\sigma^2}=0.    
\end{equation}
If we assume $r\neq0$ and $\sigma\neq\infty$, then we get

\begin{equation}
x\to x', \;\;\;\;when\;\;\;\tau\to\tau_{min}^{r=\sigma^2/2}=0, 
\end{equation}
which again shows the consistency. Here again we can predict the largest value for $\tau$ by considering the limit $\vert x-x'\vert\to\infty$. In this case we obtain 

\begin{equation}
\tau_{max}^{r=\sigma^2/2}\approx \sqrt{\frac{1}{2r}}\left(\left\vert\frac{ x-x'}{\sigma}\right\vert\right)=\frac{\vert x-x'\vert}{\sigma^2},    
\end{equation}
where we have used $r=\sigma^2/2$ for this special case. The limit $r=\sigma^2/2$ is extremely important because it corresponds to the case where the BS Hamiltonian becomes Hermitian and for this case, the information (unitarity) should be preserved in principle. We will talk more about this issue later.  

\subsection{The probability current in the non-Hermitian case: The Black-Scholes current}

We can repeat similar steps as those done in the section (\ref{CPQM}) for the non-Hermitian case. Since in finance we are normally interested in solutions of the Schr\"odinger equation which are real, then we can modify the result (\ref{derivation}) as follows

\begin{equation}
\frac{\partial \rho}{\partial t}=2C\frac{\partial C}{\partial t},
\end{equation}
where we have again used the definition $\rho=\vert C(x, t)\vert^2$. By using the Schr\"odinger equation (\ref{BSHamiltonian2}) with the Hamiltonian (\ref{BSHamiltonian}), then we get 

\begin{equation}
\frac{\partial \rho}{\partial t}=-\sigma^2C\frac{\partial^2C}{\partial x^2}+2\left(\frac{1}{2}\sigma^2-r\right)C\frac{\partial C}{\partial x}+2\vert C\vert^2
\end{equation}
The integration of the previous expression over the whole range of value of the variable $x$, gives

\begin{equation}   \label{badprob}
\frac{dP_{BS}}{dt}=2P_{BS}+\left(\frac{1}{2}\sigma^2-r\right)(C^2)^\infty_{-\infty}-\sigma^2\int^\infty_{-\infty}C\frac{\partial^2C}{\partial x^2}dx.    
\end{equation}
It is interesting to note that even if $C\to0$ asymptotically, still the probability is not conserved. This is supposed to be expected since it is a natural consequence of the non-Hermitian nature of the financial equations trying to model the Stock market. However, surprisingly even if we make the BS Hamiltonian Hermitian by using the condition $r=\frac{1}{2}\sigma^2$, it is evident from eq. (\ref{badprob}) that the probability still is not preserved in these situations. This is the case even if the last term in eq. (\ref{badprob}) vanishes after integration. In such special circumstance we have

\begin{equation}
P_{BS}=Me^{2t},    
\end{equation}
where $M$ is an integration constant. This previous equation diverges in the long term if the time flows forward, but converges if the time flows backward. The question at this point is how to solve this problem. The simplest possibility is to analytically extend the time-coordinate to the complex plane. In such a case, we simply have $t\to -i\mu$ and we will consider $\mu$ as the analytically extended time in the complex plane. In such a case, we can re-express the derivative of $\rho$ with respect to $\mu$ as  

\begin{equation}   \label{Yamamoto}   
\frac{\partial \rho}{\partial \mu}=\frac{\partial \rho_{H}}{\partial \mu}+\frac{\partial \rho_{NH}}{\partial \mu}.    
\end{equation}
Here the subindex $H$ means Hermitian and the subnidex $NH$ means non-Hermitian. The Hermitian contribution is the one defined in eq. (\ref{everytime}) with the corresponding modifications, becoming then

\begin{equation}   \label{Yamamoto2}
\frac{\partial \rho_{H}}{\partial \mu}=\frac{\partial}{\partial x}\left(i\frac{\sigma^2}{2}\left(C^*\frac{\partial C}{\partial x}-\frac{\partial C^*}{\partial x}C\right)\right)    
\end{equation}
As it is perceived by the analytically extended time $\mu$, this quantity is a conserved one with respect to the evolution of $\mu$. Note that if $r=\frac{1}{2}\sigma^2$, then $\rho_{NH}=0$ and $\rho$ is a conserved quantity in this case. Then in reality the conservation law for the probability is not explicit in the Stock market case; instead the probability is "hidden" with respect to the time $t$ and it appears with respect to its analytical extension $\mu$ when the condition $r=\frac{1}{2}\sigma^2$ is satisfied. This can be interpreted as follows: If we imagine the time-coordinate as a two-dimensional surface with one dimension corresponding to the real time $t$ and the other dimension corresponding to the imaginary time-coordinate $\mu$; then as a consequence of this, there will be two different definitions of the "energy" of the system. Then the energy surface will be a two-dimensional one and the information (probability) will be conserved in this two-dimensional surface when the Hamiltonian is Hermitian. If the Hamiltonian is non-Hermitian, then the probability will not be conserved even with respect to the analytically extended surface. In such situations, eq. (\ref{Yamamoto}) will be simplified to

\begin{equation}
\frac{\partial \rho}{\partial \mu}=\frac{\partial \rho_{NH}}{\partial \mu},   
\end{equation}
because the Hermitian part will be automatically conserved with respect to $\mu$. We can calculate the non-Hermitian component by considering the non-Hermitian contribution in eq. (\ref{BSHamiltonian}) as follows

\begin{equation}
\hat{H}_{BS}^{NH}=\left(\frac{1}{2}\sigma^2-r\right)\frac{\partial}{\partial x}.
\end{equation}
The non-Hermitian evolution of the probability can be estimated as follows

\begin{equation}
\frac{\partial \rho_{NH}}{\partial \mu}=\left(\frac{1}{2}\sigma^2-r\right)\left(\frac{\partial C^*}{\partial x}C+C^*\frac{\partial C}{\partial x}\right).    
\end{equation}

The positive sign summing the two contributions appears due to the non-Hermitian character of the derivative operator $\partial/\partial x$. In fact, $(\partial/\partial x)^+=-\partial/\partial x$ \cite{1}. We can estimate the functional dependence of the previous equation, if we consider the Kernel defined in eq. (\ref{Hammatrixagain}) which shows the functional dependence of $C(x, t)$, then we can estimate that

\begin{equation}
\frac{\partial \rho_{NH}}{\partial \mu}=2\left(\frac{\vert F\vert}{\sigma}\left(\frac{1}{2}\sigma^2-r\right)\right)^2,    
\end{equation}
where 

\begin{equation}
\vert F\vert^2=\frac{1}{2\pi\tau_\mu\sigma^2}e^{-\frac{2(x-x')}{\sigma^2}\left(r-\frac{\sigma^2}{2}\right)}.    
\end{equation}
Then the probability is not conserved in general, except when $\tau_\mu\to\infty$. Note again that if $r=\frac{\sigma^2}{2}$, then the probability is conserved. If we integrate over the whole range of prices $x$ the previous equation, we get the change in the probability as

\begin{equation}
\frac{\partial P_{NH}}{\partial \mu}=\frac{1}{2\pi\sigma^2\tau_\mu}\left(\frac{\sigma^2}{2}-r\right)e^u,  
\end{equation}
which diverges as

\begin{equation}
u=\frac{2(x-x')}{\sigma^2}\left(\frac{1}{2}\sigma^2-r\right)\to\infty,    
\end{equation}
Then it is evident that the non-conservation of probability comes out from the non-Hermitian portion of the BS Hamiltonian.

\section{The influence of the non-conservation of the probability in the prices of the Stocks}   \label{Section5}

The prices in the Stock are represented by the variable $x$. The distribution for this variable is represented by the function $C(x, t)$. The functional dependence of this wave-function is contained in in the Kernel defined in eq. (\ref{Hammatrixagain}). In this paper we have demonstrated that the information or probability in the stock market is not preserved if we consider the evolution with respect to both, the ordinary time $t$ or the analytically extended time $\mu$. The conservation of probability only appears with respect the analytically extended time $\mu$ for the case $r=\frac{\sigma^2}{2}$. This special condition appears when the non-Hermitian portion of the Hamiltonian $\hat{H}_{BS}$ vanishes. Then under the Hermiticity condition, the Kernel (\ref{Hammatrixagain}) becomes

\begin{equation}   \label{BSHHHH}
p_{BS}^{H}=e^{-r\tau}\frac{1}{\sqrt{2\pi\tau\sigma^2}}e^{-\frac{1}{2\tau\sigma^2}\left(x-x'\right)^2}.    
\end{equation}
This corresponds to a standard Gaussian distribution with center $x'$ and with time-dependent height. If we compare with eq. (\ref{Hammatrixagain}), we can notice that the effect of the non-Hermtinian portion of the BS Hamiltonian is to create a shift in the location of the center of the Gaussian. In such a case, the center of the Gaussian is also time-dependent. Under analytical extension, we can perceive from eq. (\ref{BSHHHH}) that the Kernel corresponding to the Hermitian portion of the BS Hamiltonian is just a plane-wave with a time-dependent amplitude. Under the same analytical extension, the full Kernel (\ref{Hammatrixagain}) has some additional time-dependence on the amplitude of the function. Then we can conclude that in fact, the loss of information due to the presence of the non-Hermitian contributions of the BS Hamiltonian, affect the location of the most probable price of the stock to be observed as well as the total effect of the volatility.  

\section{conclusions}   \label{Conclu}

In this paper we have analyzed the non-conservation of probability in the stock market. Since the BS Hamiltonian is non-Hermitian, it was expected that there is a loss of unitarity associated to it. This means that non-conservation of probability (or information) is expected. What is surprising is the fact that even if we eliminate the non-Hermitian portion of the Hamiltonian, still we have non-conservation of probability if it is measured with respect to the ordinary time-coordinate. However, it comes out that in these special circumstances of Hermiticity, the probability is conserved but with respect to the analytical extension of the time-coordinate as $t\to-i\mu$. Evidently, if we include the non-Hermitian portion of the Hamiltonian, then even under the analytically extended time $\mu$, loss of information (probability) is perceived. Evidently the non-conservation of the information affects the predictions of the prices in the stocks because the shape of the distribution describing the prices changes due to this effect. Our generic formalism can be extended for the analysis of more complicated financial equations. It is possible to include important concepts related to information like entropy, once the appropriate definition is established. However, in this paper we have only focused in the standard definitions corresponding to the conservation of probability in Quantum Mechanics, extended to the case of Quantum Finance.\\\\

{\bf Acknowledgement}
The authors also would like to thank the Institute of International
Business and Governance of the Open University of Hong Kong, partially funded by a grant from the Research Grants Council of the Hong Kong Special Administrative Region, China (UGC/IDS16/17), for its support.

\bibliographystyle{unsrt}  


\end{document}